\documentclass[aps,prl,twocolumn,showpacs,groupedaddress]{revtex4}

\usepackage{amsmath}
\usepackage{graphicx}
\usepackage{dcolumn}
\usepackage{bm}

\begin{document}

%\preprint{QF-HE32D}

\title{Quantum Monte Carlo study of large spin-polarized tritium clusters}

\author{I. Be\v{s}li{\'c}}
\affiliation{Faculty of Science, University of Split, HR-21000 Split,
Croatia}
\affiliation{ICAM/I2CAM, 4415 Chem Annex, UC Davis, One Shield Avenue, Davis, CA 95616}
\author{L. Vranje\v{s} Marki{\'c}}
\affiliation{Faculty of Science, University of Split, HR-21000 Split,
Croatia}
\affiliation{ Institut f\"ur Theoretische Physik, 
Johannes Kepler Universit\"at, A-4040 Linz, Austria}
\author{J. Boronat}
%\email{jordi.boronat@upc.es}
\affiliation{Departament de F\'\i sica i Enginyeria Nuclear, Campus Nord
B4-B5, Universitat Polit\`ecnica de Catalunya, E-08034 Barcelona, Spain}
\date{\today}

\begin{abstract}
This work expands recent investigations in the field of spin-polarized tritium (T$\downarrow$) clusters. We report the results for the ground state energy and structural properties of large T$\downarrow$ clusters consisting of up to 320 atoms. All calculations have been performed with variational and diffusion Monte Carlo methods, using an accurate {\it ab initio} interatomic potential. Our results for $N \leq 40$ are in good agreement with results obtained by other groups. Using a liquid-drop expression for the energy per particle, we estimate the liquid equilibrium density, which is in good agreement with our recently obtained results for bulk T$\downarrow$. In addition, the calculations of the energy for large clusters have allowed for an estimation of the surface tension. From the mean-square radius of the drop, determined using unbiased estimators, we determine the dependence of the radii on the size of the cluster and extract the unit radius of the T$\downarrow$ liquid.  
\end{abstract}


\maketitle

\section{Introduction}\
The extreme quantum nature of electron spin-polarized hydrogen (H$\downarrow$) and its isotopes, spin-polarized deuterium (D$\downarrow$) and spin-polarized tritium (T$\downarrow$), promoted renewed theoretical interest  \cite{blume1,hydrogen,tritium1,smallhyd,tritiumHD,tritium} in their condensed phases. They are characterized by the small mass of their atoms and their weakly attractive interatomic  potential, which is  very accurately determined in {\em ab initio} calculations. Already in the seventies, several theoretical predictions of the spin-polarized hydrogen systems  appeared \cite {etters,miller}.
In 1976, Stwally and Nosanow~\cite{StwaleyNosanow} underlined H$\downarrow$ as a first candidate for achieving a Bose-Einstein condensate (BEC) state. Their theoretical prediction was experimentally confirmed in 1998 by Fried {\em et al.}~\cite{fried}, who succeeded to overcome demanding experimental obstacles and reported the formation of a BEC state with H$\downarrow$ atoms.   

Further investigation related to possible candidates for BEC state in hydrogen systems was done by Blume {\em et al.}. In their work~\cite {blume1},  spin-polarized tritium clusters as well as optically pumped tritium condensate were theoretically investigated for the first time. Blume {\em et al.} reported  their DMC results for the ground state energy and structural  properties of T$\downarrow$ clusters consisting of up to  40 atoms. In addition, it was also shown that the smallest T$\downarrow$ cluster is a trimer, i.e. (T$\downarrow$)$_3$; negative ground-state energy was not  obtained for the dimer (T$\downarrow$)$_2$, which means that T$\downarrow$ trimer is an example of Borromean or halo state. The same conclusion for (T$\downarrow$)$_3$, obtained with the finite element method, was  reported by Salci {\em et al}~\cite{salci}. Because of evident resemblance of bosonic T$\downarrow$ and $^4$He atoms Blume {\em et al.}~\cite{blume1} also compared general properties of both types of clusters. They  showed that common attributes of T$\downarrow$ clusters are weaker binding and greater interparticle distances between atoms, in comparison with  $^4$He clusters having the same number of atoms. In the same work, results of coupled-channel scattering calculations for two T$\downarrow$ atoms are reported, indicating the possibility for formation of a tritium condensate using its broad Feshbach resonance. 

Mixed clusters consisting of spin-polarized hydrogen-tritium and deuterium-tritium atoms have also been investigated.~\cite{smallhyd,tritiumHD} It has been shown that three T$\downarrow$ atoms are needed to bind one D$\downarrow$ atom in a stable system, making thus clusters (T$\downarrow$)$_N$D$\downarrow$ stable  for all $N\geq 3$. On the contrary, it has been shown that even 60 T$\downarrow$ are not enough to bind one H$\downarrow$ in a stable system. Namely, the ground-state energy of the cluster (T$\downarrow$)$_{60}$H$\downarrow$ is within the error bar equal to the ground state energy of  (T$\downarrow$)$_{60}$, leading to the conclusion that clusters (T$\downarrow$)$_N$H$\downarrow$ for $N\leq 60$ are effectively unstable or are at the threshold of binding. Due to the more complicated calculations in the case of several fermionic D$\downarrow$ atoms, so far only the stability limits of small mixed spin-polarized deuterium-tritium clusters having up to 5 D$\downarrow$ atoms have  been examined.~\cite{smallhyd} 

Despite  the lack of experimental verification, bulk properties of all spin-polarized hydrogen isotopes have been theoretically predicted. Bulk properties of the D$\downarrow$ system are conditioned by the number of occupied nuclear spin states.~\cite{panoff,flynn,skjetne} If only one nuclear spin state is occupied (D$\downarrow$$_{1}$), the system is in a gas state at zero pressure, while in the case of two (D$\downarrow$$_{2}$) or three (D$\downarrow$$_{3}$) equally occupied nuclear spin states, the system remains liquid at zero pressure and zero temperature.
Extensive investigations of the H$\downarrow$ and T$\downarrow$ bulk systems have been carried out recently with the diffusion Monte Carlo (DMC) method~\cite{hydrogen,tritium}, which provides exact results within errorbars for bosonic systems. Using the DMC method for bulk H$\downarrow$ and T$\downarrow$, the energy per particle, structural properties, as well as densities and the pressure of the gas (liquid)-solid transition have been predicted. An accurate calculation of the ground-state energy per particle in bulk H$\downarrow$ has allowed the confirmation of its gas nature in the limit of zero temperature and up to 170 bar, point at which H$\downarrow$ solidifies. A similar investigation of bulk T$\downarrow$ has revealed that the system is a liquid up to 9 bars, where it crystallizes.~\cite{tritium}  

In this work, we expand previously reported studies of pure T$\downarrow$ clusters.\cite{blume1,smallhyd,tritiumHD} We report the ground-state energy of clusters having up to 320 atoms, as well as their structural properties, obtained with the DMC method. From the density profiles, we estimate the thickness of the clusters' surface. Justification for carrying out demanding calculations for large clusters lies in the fact that the present results for clusters can be used to extrapolate precise equilibrium T$\downarrow$ bulk properties. A goal of our investigation is to examine the validity of the liquid-drop formulas when they are applied to T$\downarrow$ clusters, as it was done in the past for $^3$He and $^4$He clusters.~\cite {pandharipande,chin} In helium clusters, liquid-drop formulas were successfully applied and the results for the equilibrium energy per particle and the unit liquid radius were in good agreement with experimental studies.~\cite {pandharipande,chin} We have used the energy per particle of the T$\downarrow$ clusters to extrapolate the equilibrium energy per particle in bulk T$\downarrow$. We compare the estimated result  with the energy per particle calculated in a recent DMC study of a bulk T$\downarrow$.~\cite{tritium} In addition, the surface tension of liquid T$\downarrow$ is estimated and compared with known results for $^3$He and $^4$He liquids. Furthermore,  we extract the unit radius of the liquid using the average distance of the particles to the centre of mass of the cluster. 

In Sec. II, we report briefly the DMC method and discuss the trial wave
functions used for importance sampling of the clusters.  Sec. III reports the results obtained by the DMC simulations. Finally, Sec. IV comprises a summary of the
work and an account of the main conclusions.

\section{Method}
\label{method}

The starting point of the DMC method is the Schr\"odinger equation 
written in imaginary time,
\begin{equation}
-\hbar \frac{\partial \Psi(\bm{R},t)}{\partial t} = (H- E_{\text r}) \Psi(\bm{R},t) \ ,
\label{srodin}
\end{equation}
where $E_{\text r}$ is a constant acting as a reference energy and
$\bm{R} \equiv (\bm{r}_1,\ldots,\bm{r}_N)$ collectively denotes particle positions.

The $N$-particle Hamiltonian, $H$, is given as
\begin{equation}
 H = -\frac{\hbar^2}{2m} \sum_{i=1}^{N} \bm{\nabla}_i^2 + \sum_{i<j}^{N} V(r_{ij})  \ ,
\label{hamilto}
\end{equation}
where $V(r)$ is the interaction potential. The interatomic interaction between tritium atoms is described with the
spin-independent central triplet pair potential $b$$^3\Sigma_u^+$, which was  
determined in an essentially 
exact way by Kolos and Wolniewicz~\cite{kolos}. 
As in our recent DMC calculations of bulk H$\downarrow$ and T$\downarrow$,\cite{hydrogen,tritium} we have  used the recent extention of Kolos and Wolniewicz data to larger interparticle distances by Jamieson \textit{et al.}
(JDW).~\cite{jamieson} The potential is finally constructed using a cubic spline interpolation of JDW data, which is smoothly connected to the long-range behavior of the T$\downarrow$-T$\downarrow$  potential as
calculated by Yan \textit{et al.}.~\cite{yan}
The JDW potential used in the present work has a core diameter
$\sigma=3.67$~\AA\, and a minimum of $-6.49$ K at a distance $4.14$ \AA.  We have previously verified that the addition of mass-dependent adiabatic corrections (as calculated by Kolos and Rychlewski~\cite{kolos2}) to the JDW potential does not change the energy of bulk spin-polarized tritium.~\cite{tritium} It is worth mentioning that within the Born-Oppenheimer approximation it has been explicitly shown that in the spin-aligned electronic state, tritium 
nuclei behave as effective bosons.~\cite{freed}

DMC solves stochastically the Schr\"odinger
equation (\ref{srodin}) by multiplying $\Psi(\bm{R},t)$ with the $\psi(\bm{R})$, a trial wave function used for importance sampling, and rewriting Eq. (\ref{srodin}) in terms of the mixed distribution $\Phi(\bm{R},t)=
\Psi(\bm{R},t) \psi(\bm{R})$. Within the Monte Carlo framework, $\Phi(\bm{R},t)$ is represented by a set of \textit{walkers}. 
In the limit $t \rightarrow \infty$ only the lowest energy 
eigenfunction, not
orthogonal to $\psi(\bm{R})$, survives and then the sampling of the ground
state is effectively achieved. Apart from statistical uncertainties, the
energy of a $N$-body bosonic system is exactly calculated.

In the present simulations Jastrow trial wave functions have been used, 
\begin{equation}
\psi_{\text J}(\bm{R}) = \prod_{i<j}^{N} f(r_{ij}),
\label{trial}
\end{equation}
with a two-body correlation function $f(r)$, 
\begin{equation}
f(r_{ij})=\exp\left[-\left(\frac{b}{r_{ij}}\right)^{5}-sr_{ij}^n\right] \ ,
\label{eq:trial1}
\end{equation}
where $n=1$ or $n=2$, depending on the size of the cluster, and $b$ and $s$ are variational parameters.
Previous experience in work with small pure and mixed spin-polarized tritium clusters ~\cite{smallhyd} has shown that the best choice for the two-body correlation function is obtained with $n=1$. 
We have used this type of function for clusters having $N\leq 60$ atoms, but for larger clusters the variational energies obtained with this type of function worsen significantly. Then, for larger clusters, it is better to consider the model with $n=2$, which is similar to $f(r)$ which has been used in recent investigations of vortices in large $^4$He clusters.~\cite{bighel}  
Eq. (\ref{eq:trial1}) with $n=2$ defines much better the confinement of space in which large number of T$\downarrow$ atoms is settled and definitely provides better VMC energies for clusters having $N\geq 80$ atoms. 

The optimization of the trial wave functions has been done for all clusters by means of the  variational Monte Carlo method. For clusters with $N\leq 60$ the best variational parameters vary from $b$=3.574 \AA ~to $b$=3.605 \AA~ and from $s$=0.0328 \AA$^{-1}$ to $s$=0.0073 \AA$^{-1}$ for increasing $N$.
In case of clusters having $N\geq 80$ atoms, the parameter  $b$  assumes values from 3.574 \AA ~to 3.605 \AA, while at the same time $s$ varies from 000162 \AA$^{-2}$ to 0.0000145 \AA$^{-2}$. It is worth noticing that in both types of two-body correlation functions ($n=1,2$) the parameter $b$ remains practically constant, while $s$ always decreases with $N$.

For several clusters we have verified that 1000 walkers are enough for excluding the bias coming from the size of the population ensemble used in a simulation. Thus,  we have decided to employ this number of walkers in all the remaining DMC calculations. The same conclusion emerged also from previous experience in pure and mixed T$\downarrow$ clusters.

In order to eliminate bias coming from the time-step value used in simulations, all calculations have been performed with several $\Delta t$ time-steps which assume values within the interval $5\times 10^{-4} - 1.3\times 10^{-3}$K$^{-1}$. From the obtained results, we have extrapolated the result  to $\Delta t \rightarrow 0$. In accordance with the DMC method used in this work, which is accurate to second order in the time step~\cite {boro}, the extrapolation is made with a quadratic function. Second order DMC enables the use of greater time-steps than the linear DMC method.

\section{Results}
\begin{figure}[t]
\centering
 \includegraphics[width=0.4 \textwidth]{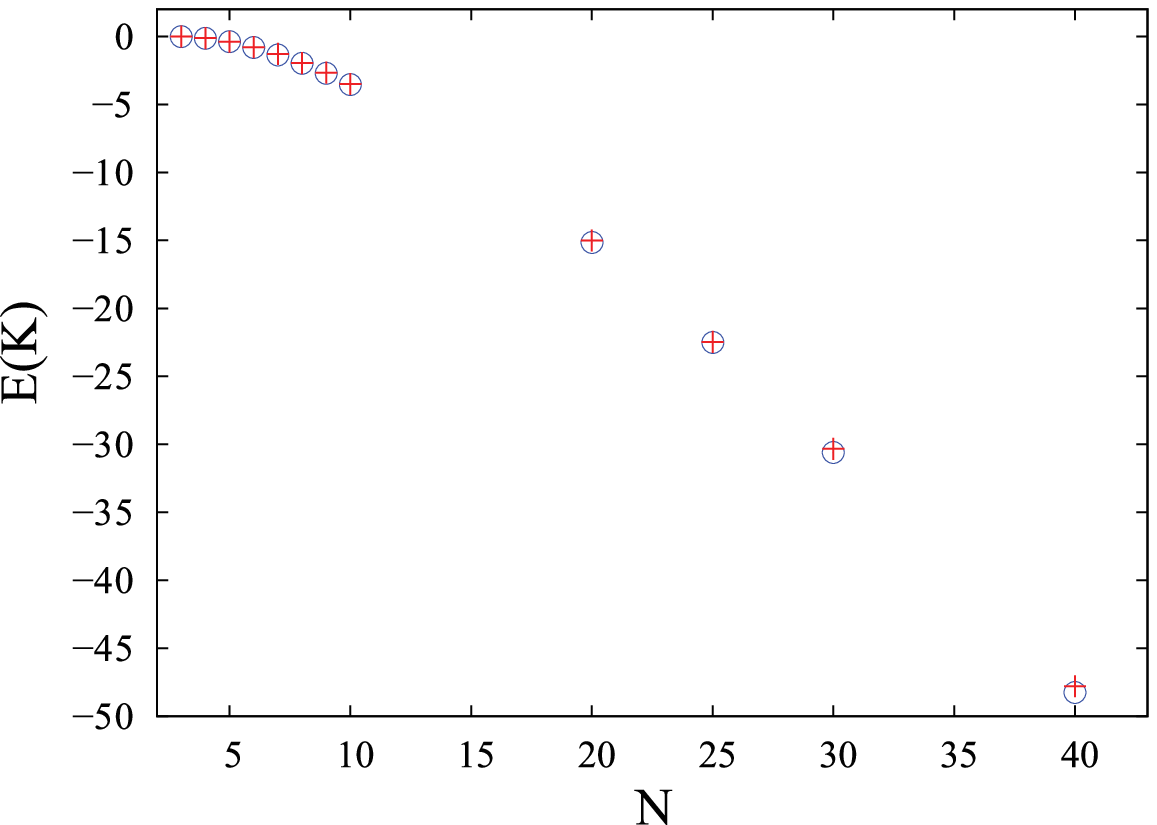}
\\
 \caption{I. Be\v{s}li{\'c} {\it et al.}}
     \label{fig:comparison}	
\end{figure}

Our DMC results for the ground-state energy per particle and radii of the investigated clusters are given in Table \ref{tab:energy}. 
For clusters consisting of up to 10 T$\downarrow$ atoms we have already shown~\cite{smallhyd} good agreement with results obtained by Blume et {\it al.} \cite {blume1}. Here, we extend this comparison for clusters consisting of up to 40  T$\downarrow$  atoms. In Fig. \ref{fig:comparison}, comparison of our results and the ones by Blume {\it et al.} is shown. As in the case of small clusters, we report slightly lower ground-state energy for all clusters (T$\downarrow$)$_N$. These differences are mainly due to the fact that there are  small differences in the potential of interaction employed in two simulations. Namely, Blume \textit{et al.}~\cite {blume1} have included in the Hamiltonian the damped three-body Axilrod-Teller potential term \cite{threebody} which causes a slight raise of the ground-state energy.

In Ref. \onlinecite {blume1}, weaker binding and greater spread in the T$\downarrow$ clusters were emphasized as the main difference between small (T$\downarrow$)$_N$ and ($^4$He)$_N$ clusters, for the same $N$ and up to 40.  Similar comparison can be done for large clusters. Namely, the energy per particle of the largest investigated $^4$He cluster with the DMC method, ($^4$He)$_{112}$,  is -3.780(3) K \cite {whaley}. This can be compared with the energy per particle of the largest  T$\downarrow$ cluster, (T$\downarrow$)$_{320}$, which is -2.286(8) K (Table \ref{tab:energy}). It is clear from these results that in the ($^4$He)$_{112}$ cluster, which consists of almost three times smaller number of atoms than (T$\downarrow$)$_{320}$ cluster, binding is significantly stronger. From that, we conclude that the binding  in large T$\downarrow$ clusters is very weak, as it was already concluded for small clusters \cite {blume1}.

The energy per particle of quantum liquid clusters as a function of $N$ is well reproduced by a liquid-drop model,~\cite {pandharipande,chin}
\begin{equation}
E(N)/N=E_{v}+xE_{s}+x^{2}E_{c}  \ ,
\label{eq:liquiddrop}
\end{equation}
where $E_v$, $E_s$ and $E_c$ are respectively the volume, surface and curvature terms, and the  variable $x$ is defined as $x$=$N^{-1/3}$.

\begin{figure}[t]
\centering
 \includegraphics[width=0.4 \textwidth]{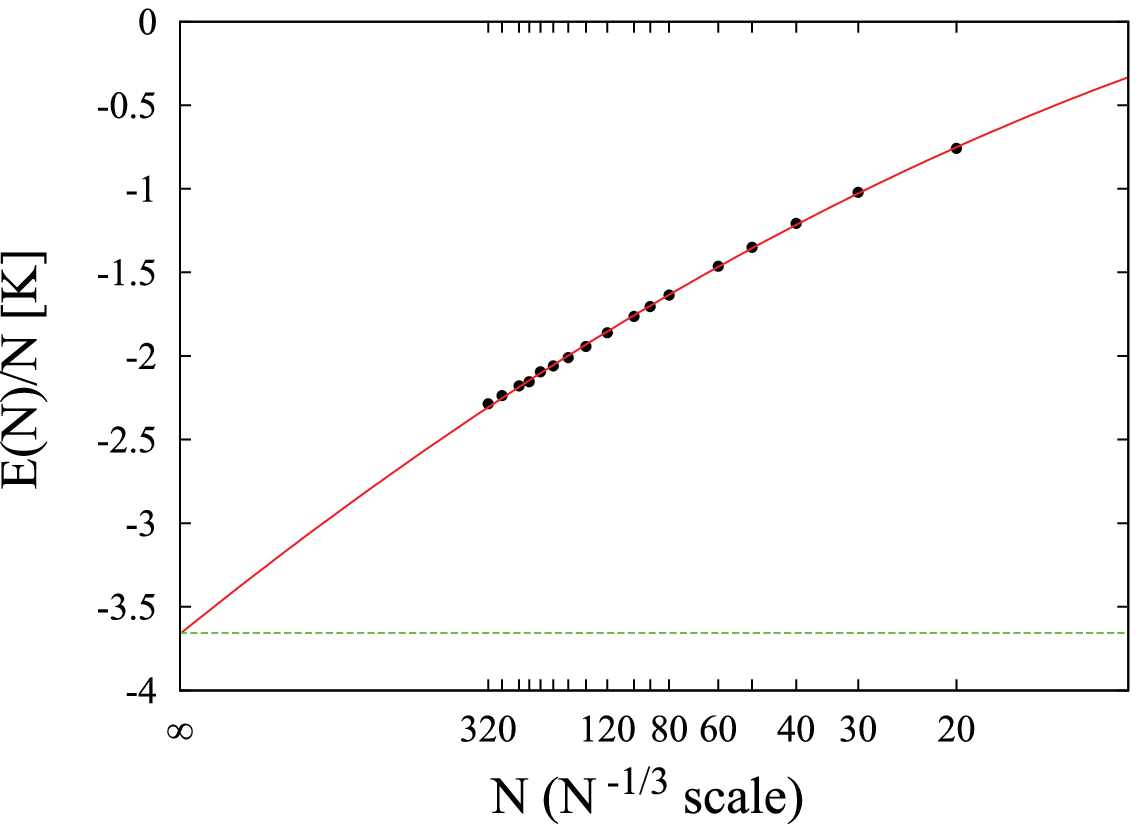}
\\
 \caption{I. Be\v{s}li{\'c} {\it et al.}}
     \label{fig:energyfit}	
\end{figure}

In Fig. \ref{fig:energyfit}, we have plotted results for the energy per particle from Table \ref{tab:energy}, as well as a line on the top of data, which represents the best fit with expression (\ref{eq:liquiddrop}). We have included in our fit all the investigated clusters and the best set of parameters obtained with the above mentioned fit is: $E_v$= -3.66(3) K, $E_s$= 10.2(2) K and $E_c$= -6.1(4) K. 

The parameter $E_v$ represents the energy per particle of bulk liquid T$\downarrow$ at the equilibrium density. This extrapolated result is in a very good agreement with our recent results obtained in calculations of bulk T$\downarrow$ \cite {tritium}. With the DMC method we obtained $\rho_0=0.007466(7) $ \AA$^{-3}$ as the equilibrium density of liquid T$\downarrow$ and $e_0=-3.656(4)$ K as the energy per particle at that density.

It is also important to emphasize that the parameters $E_v$, $E_s$ and $E_c$ have been obtained without including in the fit the bulk energy per particle at equilibrium. Furthermore, the parameters in Eq. (\ref{eq:liquiddrop}) remain practically the same when the energy per particle at the equilibrium density is included in the fit. Thus, the demanding DMC calculations have been worthwhile because the result of our liquid-drop model (\ref{eq:liquiddrop})  does not depend on the knowledge of the equilibrium energy per particle of the bulk. Also, we have determined that (T$\downarrow$)$_{280}$  is the 'smallest' cluster needed to be  included in the fit in order to extrapolate the parameter $E_v$ properly. Namely, using just DMC results for clusters having $N=20-280$ atoms in the fit we get $E_v$= -3.69(3) K, which is within the error bars the same as the result obtained including the largest cluster ($N$=320, $E_v$=-3.66(3) K). However, the extrapolation of the equilibrium energy per particle with fits including just results for clusters smaller than (T$\downarrow$)$_{280}$ always produces lower energy than the one calculated for the bulk.  For example, the extrapolated bulk energy with results for clusters having up to 120 atoms is -3.85(2) K, which is around 5\% lower than $e_0$. Similar conclusions about this fit emerged for $^4$He clusters \cite {chin}, where it was emphasized that an accurate extrapolation of the bulk equilibrium energy from finite cluster calculations should include relatively large clusters. 

We have also tried to fit all the obtained data for $E/N$ with a linear function in $N^{-1/3}$, as in Ref. \onlinecite {chin}, but that kind of fit has not been so precise as the one performed including a quadratic dependence on $N^{-1/3}$. With the linear fit $E_v$= -3.26(3) K, which is around 11\% higher than the $e_0$ obtained for bulk, showing the necessity of including the second-order term.

The second parameter $E_s$ extracted from Eq. (\ref{eq:liquiddrop}) is related to the surface tension of liquid T$\downarrow$ through
\begin{equation}
t=\frac{E_{s}}{4{\pi}r_{0}^{2}} \ ,
\label{eq:surten}
\end{equation}
where $r_0$ is the unit radius of the liquid.
The unit radius of the liquid can be determined in two ways, using the result of the equilibrium density of the bulk liquid 
\begin{equation}
\label{eq:radius}
\frac{4{\pi}}{3}r_{0}^{3}{\rho}_{0}=1 \ ,
\end{equation}  
or from the expression
\begin{equation}
\label{eq:clusterrad}
{r_{0}(N)}= \left[\frac{5}{3}\langle r^{2}(N) \rangle\right]^{1/2} N ^{-1/3} \ ,
\end{equation}
where $\langle r^{2}(N) \rangle$ is the mean-square radius of a cluster with $N$ atoms.

Using the result for the equilibrium density of the T$\downarrow$ liquid $\rho_0=0.007466(7)$ \AA$^{-3}$ \cite{tritium} and Eq. (\ref{eq:radius}), we obtain $r_0=$3.18(1) \AA~.

\begin{figure}[t]
\centering
 \includegraphics[width=0.4 \textwidth]{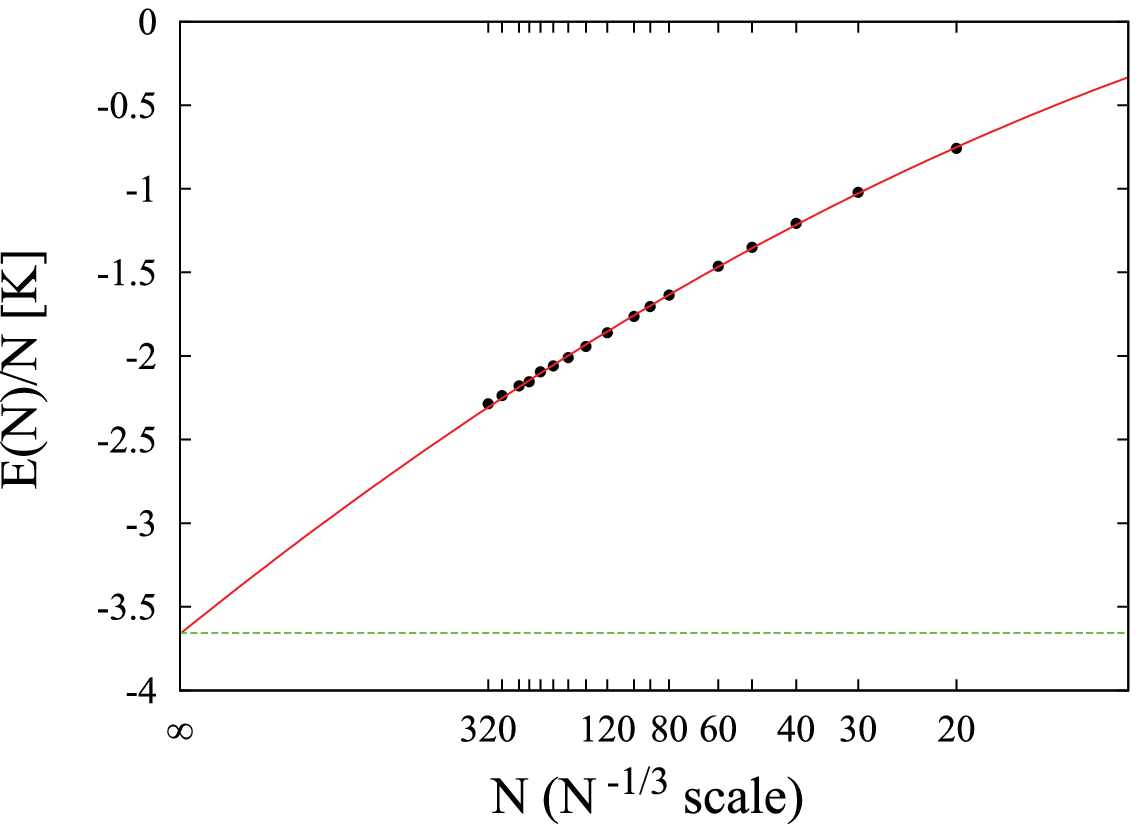}
\\
 \caption{I. Be\v{s}li{\'c} {\it et al.}}
     \label{fig:radiusfit}	
\end{figure}

The second method for obtaining the unit radius, Eq. (\ref{eq:clusterrad}), was previously used in the study of $^4$He clusters.~\cite {pandharipande,chin}
In Ref. \onlinecite {pandharipande}, it is emphasized that only those $^4$He clusters with more than ten atoms have a radius which increases approximately as $N ^{1/3}$. Since in our calculations we obtain unbiased mean-square radii of clusters with pure estimators, we tried to interpolate our data for clusters radii, $R_{cm}(N)=\sqrt{\langle r^{2}(N) \rangle}$, with several polynomial functions of the variable $x=N^{1/3}$. As in the case of the interpolation of the energy per particle, we have included all clusters having $N=20-320$ atoms. In Fig. \ref{fig:radiusfit}, clusters' radii obtained from the calculations are plotted, as well as two lines on top of data which represent two interpolations, using functions
\begin{equation}
{g_{1}(x)}=a+bx  \\,
\label{eq:radfit1}
\end{equation}
and
\begin{equation}
{g_{2}(x)}=cx +\frac{d}{x} \\.
\label{eq:radfit2}
\end{equation}
The parameters extracted from the fits are: $a=1.85(12)$\AA, $b=2.34(3)$\AA, $c=2.55(1)$\AA\, and $d=3.73(12)$\AA.   
Using these interpolation parameters and the definition of unit radii given in (\ref{eq:clusterrad}), in the limit $N\rightarrow \infty$, we extract an equilibrium radius $r_0=3.02(4)$ \AA~ using the function (\ref{eq:radfit1}) and $r_0=3.29(1)$ \AA~ using the function (\ref{eq:radfit2}). 
Since in both cases the quality of the fit is very good, we cannot state which of the two extracted results for equilibrium unit radius should be considered as the better estimation. We can thus only conclude that the equilibrium unit radius assumes a value within  the interval from 3.02(4) \AA~ to 3.29(1) \AA.  

Therefore, the estimation of the equilibrium unit radius from the results obtained in clusters calculations is very sensitive to the choice of interpolating function. Because of that, we decided to include the unit radius 3.18(1) \AA~, derived from the bulk T$\downarrow$, in the estimation of the surface tension.

It is useful to compare the value $r_0$ with $\sigma=3.67$ \AA. If we consider $r_0$ as the radius of the sphere that one T$\downarrow$ atom occupies at equilibrium density in a liquid, $\sigma$ has to be smaller than  2 $r_0$, as it is in our case.
On the other hand, the value of $r_0$ of liquid T$\downarrow$ is also greater than the unit  radius of liquid $^4$He ($r_0$=2.1799 \AA\,) \cite {chin}. This also explains the greater spread in T$\downarrow$ clusters because it is clear that T$\downarrow$ atoms occupy more space than $^4$He atoms at equilibrium density.

From $r_0$, it is possible to calculate the liquid surface tension $t$ using expression (\ref{eq:surten});  we have obtained $t=0.08$ K\AA$^{-2}$. There is no experimental result for the surface tension of liquid T$\downarrow$, and this prediction is, to the best of our knowledge, the first estimation of $t$ for liquid T$\downarrow$.
Contrary to liquid T$\downarrow$, the surface tensions of $^4$He and $^3$He liquids have been experimentally investigated and the measured values are respectively 0.27 K\AA$^{-2}$ and 0.11 K\AA$^{-2}$.~\cite {pandharipande}  In the case of liquid T$\downarrow$ our estimated value of the surface tension is even smaller than the surface tension of the $^3$He liquid, although bulk T$\downarrow$ is a bosonic system. Explanation for such a small value of the surface tension lies in the fact that the interaction between T$\downarrow$ atoms is described with a very shallow potential.

In addition to the ground-state energy, we have also studied the structure of T$\downarrow$ clusters.  Exact estimators of DMC method have been employed to calculate values such as the pair distribution function $P(r)$, as well as the distribution of particles with respect to the centre of mass of the cluster  $\rho(r)$. Possible bias in our results coming from the type of trial wave function used in the simulations is resolved with the use of pure estimators which ensure  unbiased results.~\cite{pures}  

\begin{figure}[t]
\centering
 \includegraphics[width=0.4 \textwidth]{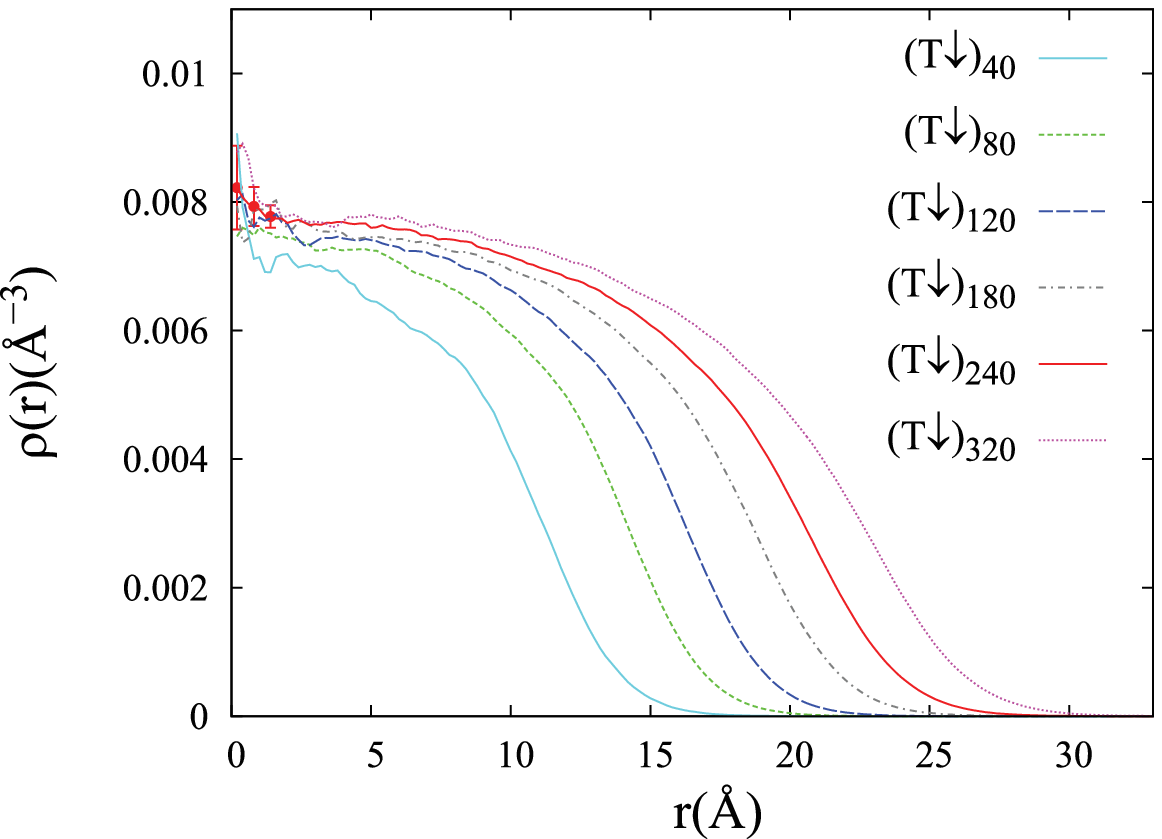}
\\
 \caption{I. Be\v{s}li{\'c} {\it et al.}}
     \label{fig:density}	
\end{figure}

In Fig. \ref{fig:density}, the density distributions of T$\downarrow$ clusters having 40, 80, 120, 180, 240 and 320 atoms are plotted. The density profiles show that the cluster size grows when the number of atoms in the cluster increases, and that the central densities of the  largest clusters are very similar to the bulk equilibrium density $\rho_0=0.007466(7) $ \AA$^{-3}$.

\begin{figure}[t]
\centering
 \includegraphics[width=0.4 \textwidth]{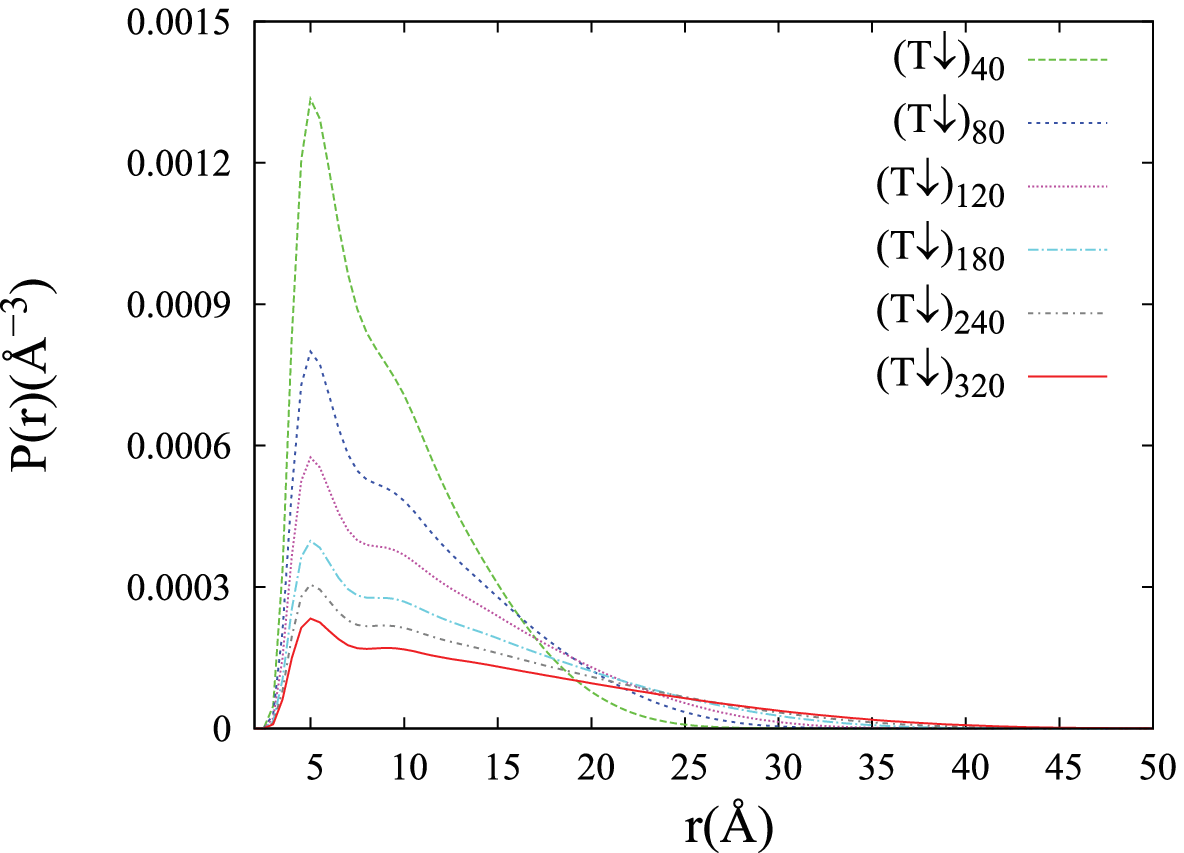}
\\
 \caption{I. Be\v{s}li{\'c} {\it et al.} }
     \label{fig:interparticle}	
\end{figure}

The increase of cluster size with increasing number of atoms can also be seen from the pair distribution  function $P(r)$ shown in Fig. \ref {fig:interparticle} for the same clusters. $P(r)$ is normalized such that $\int P(r)r^2dr=1$. A significant decay of the peak height for the largest clusters, as well as the growing probability for larger interparticle distances in large clusters, is a clear evidence of the size spreading tendency.

The surface thickness of clusters $s_t$ can be estimated from the density profiles as a difference of radii at which the central density $\rho_c$=$\rho(r=0)$ has decreased from 90\% to 10\% of its value. From the plotted density profiles in Fig. \ref{fig:density} it is obvious that the density error bars are large for small distances. In order to determine the central density as precisely as possible we have tried to fit the density profile with the function used in Ref. \onlinecite{densityfit}
\begin{equation}
{\rho(r)}=\frac{\rho_0}{(1+e^{\beta(r-r_0)})^\delta}  \\~,
\label{eq:fitden}
\end{equation}
where $\rho_0$, $\beta$, $r_0$ and $\delta$ are fitting parameters and $r$ is the distance to the centre of mass of the cluster.
We find that for T$\downarrow$ clusters Eq. (\ref{eq:fitden}) can be employed to model density profiles of small clusters, while for greater clusters the same model reproduces poorly the calculated density profiles at small distances. Since the small distances are important for our calculation, we have decided to fit the calculated density profiles to a constant function for distances up to some value $r_1$. We have varied the value of $r_1$ from 2 \AA\, to 4 \AA, increasing it with the growing size of the cluster. We  have considered the constant obtained with the fit as a central density value and used it in further estimation of the clusters' surface thickness. The results for the surface thickness are reported in Table \ref{tab:energy}. We can compare our results with the surface thickness of $^4$He clusters \cite{pandharipande,stringari}. Using the VMC method Pandharipande {\em et al.}~\cite{pandharipande} showed that in $^4$He clusters the surface thickness is $\sim$7\AA~for clusters $N\geq 112$.
In the case of T$\downarrow$ clusters, for $N\geq 100$, the surface thickness is significantly greater than in $^4$He 
(Table \ref{tab:energy}). This is expected due to the evidently greater interparticle distances in T$\downarrow$ clusters, which is a direct consequence of the shallow attractive part of T$\downarrow$-T$\downarrow$ interaction potential. With the density functional approach, Stringari {\em et al.}~\cite{stringari} calculated the surface thickness of  several $^4$He clusters and we can compare those results with our results for T$\downarrow$ clusters having 20, 40 and 240 atoms. The reported surface thickness for clusters $^4$He$_{20}$, $^4$He$_{40}$ and $^4$He$_{240}$ are respectively 8.8 \AA, 9.0 \AA~and 9.3\AA. A comparison with (T$\downarrow$)$_N$ reveals larger surface thickness of $^4$He$_N$ for $N$=20,40 and smaller surface thickness for $N$=240. Also, it can be noticed that the surface thickness reported by Stringari {\em et al.} is not a linear function of the number of atoms. Contrary, in the case of T$\downarrow$ clusters, we observe that the surface thickness is almost a linear function of $N$, up to $N$= 320 atoms. We have tried to predict the surface thickness of clusters having more than 320 atoms by fitting our data with the function used in Ref. \onlinecite{densityfit} to predict the width of a free surface. However, with the present results we have not been able to determine the asymptotic value of the surface thickness.  Saturation should be probably seen with results for clusters having more than $N$=320 atoms, but the DMC calculations are already difficult with 320 atoms.

\section{conclusions}
General characteristics of large spin-polarized tritium T$\downarrow$ clusters have been investigated using the DMC approach. The ground-state energies of clusters consisting of up to 40 T$\downarrow$ atoms have been compared with previously published results.  For clusters having more than 40 atoms the ground-state energies, as well as the structure description, are determined for the first time. This prediction relies on the use of a very precise potential of interaction between T$\downarrow$-T$\downarrow$ atoms. The present results for the ground-state clusters' energy are also used to extract the energy per particle of liquid T$\downarrow$ at equilibrium density using a liquid-drop model. The extrapolation using a liquid-drop formula gives a value $E_v=-3.66(3)$ K for the energy per particle in the equilibrium bulk system, which is in very good agreement with the result from a recent DMC calculation of the bulk, $e_0=-3.656(4) K$~\cite{tritium}. 

The radii of clusters are calculated with pure estimators and those results are used to estimate the interval in which the unit radius of the liquid is expected.  The result for the unit radius from the bulk calculation lies in the estimated interval. The latter value of the bulk unit radius has been employed to estimate the surface tension $t$ of bulk T$\downarrow$, $t$=0.08 K\AA$^{-2}$. In addition, the surface thickness of clusters has been estimated from the clusters' density profiles. 

As it is already shown for clusters consisting of up to 40 T$\downarrow$ atoms~\cite{blume1}, it is concluded that large spin-polarized tritium clusters are less bound and are more diluted than $^4$He clusters with the same number of atoms, i.e., interparticle distances are significantly greater in the corresponding T$\downarrow$ clusters.

\acknowledgments

J. B. acknowledges support from DGI (Spain) Grant No.
FIS2005-04181 and Generalitat de Catalunya Grant No. 2008SGR-04403. I.B and L.V.M. acknowledge
support from MSES (Croatia) under Grant No. 177-1770508-0493. I. B. also acknowledges support from
L'Or\'eal ADRIA d.o.o. and Croatian commission for Unesco, as well as from U.S. National Science Foundation I2CAM International Materials Institute Award, Grant DMR-0645461.
We also acknowledge the support of the Central Computing Services at
the Johannes Kepler University in Linz, where part of the computations was
performed.  In addition, the resources of the Isabella
cluster at Zagreb University Computing Centre (Srce) and Croatian National
Grid Infrastructure (CRO NGI) were used.

\newpage

\begin{table}[t!]
\begin{center} 
\begin{tabular}{cccc} 
\\ \hline
\hline
~~$N$~~&~~~~~~~$E$((T$\downarrow$)$_N$)/$N$~~~~~~~~ &~~~~~~~ $R_{cm}(N) $ ~~~~~~~ &~~~~~~~ $s_t$ ~~~~~~~
\\ \hline
20    & -0.758 (0.004) &  8.4 (0.4) &  7.4(0.3) \\
30    & -1.020 (0.005) &  9.1 (0.4) &  7.6(0.3) \\
40    & -1.206 (0.004) &  9.8 (0.5) &  8.0(0.3) \\
50    & -1.350 (0.005) & 10.3 (0.5) &  8.0(0.3) \\ 
60    & -1.464 (0.004) & 10.9 (0.5) &  8.4(0.3) \\  
80    & -1.635 (0.004) & 11.8 (0.9) &  8.6(0.3) \\
90    & -1.704 (0.004) & 12.2 (0.9) &  8.8(0.3) \\
100   & -1.763 (0.006) & 12.6 (0.9) &  9.0(0.3) \\
120   & -1.861 (0.006) & 13.3 (0.9) &  9.4(0.3) \\ 
140   & -1.943 (0.005) & 13.9 (0.9) &  9.8(0.3) \\
160   & -2.009 (0.006) & 14.5 (0.9) & 10.6(0.3) \\
180   & -2.059 (0.014) & 15.1 (0.9) & 11.0(0.3) \\
200   & -2.095 (0.008) & 15.7 (0.9) & 11.2(0.3) \\
220   & -2.154 (0.009) & 16.0 (1.0) & 11.6(0.3) \\
240   & -2.179 (0.014) & 16.6 (1.0) & 12.2(0.3) \\
280   & -2.237 (0.006) & 17.4 (1.1) & 13.4(0.3) \\
320   & -2.286 (0.008) & 18.2 (1.1) & 13.0(0.3) \\
\hline
\hline
\end{tabular} 
\end{center}
\caption{Energy per particle (in K), radii and surface thickness (in \AA ) of investigated T$\downarrow$ clusters.}
\label{tab:energy}
\end{table}

%\clearpage

\textbf{Figure captions}\\

FIG. 1: Comparison of calculated ground-state energies of clusters (T$\downarrow$)$_N$ for $N\leq $ 40 atoms (circles) with the results reported by Blume \textit{et al.} in Ref. \onlinecite {blume1} (crosses). The error bars of the DMC energies are smaller than the size of the symbols.
	
FIG. 2: Energy per particle $E(N)/N$ for (T$\downarrow$)$_N$ clusters reported in Table \ref{tab:energy}. Given abscissa is $N$ on an $N^{-1/3}$ scale. The bulk value obtained in \cite {tritium} is plotted with a dashed line.
	
FIG. 3: Radii of (T$\downarrow$)$_N$ clusters. The abscissa is $N$ on an $N^{1/3}$ scale. The interpolation function (\ref{eq:radfit1}) is displayed with a solid line and the function (\ref{eq:radfit2}) with a dashed line.

FIG. 4: Density profiles for several T$\downarrow$ clusters. Errorbars are large for small distances to the centre of mass of the clusters, as indicated in the figure for the cluster having 240 atoms, and decrease for larger distances. 
	
FIG. 5: Pair distribution function for several T$\downarrow$ clusters.
	
%\newpage

%\clearpage

%\clearpage

%\clearpage

%\clearpage

\end{document}